%% file: Camera-ready-arxiv.tex
\documentclass[letterpaper]{article} 
\usepackage{aaai23}  
\usepackage{times}  
\usepackage{helvet}  
\usepackage{courier}  
\usepackage[hyphens]{url}  
\usepackage{graphicx} 
\usepackage{natbib}  
\usepackage{caption} 
\usepackage{algorithm}
\usepackage{algorithmic}

\usepackage{newfloat}
\usepackage{listings}
\DeclareCaptionStyle{ruled}{labelfont=normalfont,labelsep=colon,strut=off} 
\floatstyle{ruled}
\newfloat{listing}{tb}{lst}{}
\floatname{listing}{Listing}
\usepackage[switch]{lineno}
\usepackage{booktabs}
\usepackage{multirow}
\usepackage{amsmath, bm}
\usepackage{epstopdf}
\input{command}

\title{DeAR: A Deep-Learning-Based Audio Re-recording Resilient Watermarking}
\author{
	Chang Liu\textsuperscript{\rm 1}, 
	Jie Zhang\textsuperscript{\rm 1,2},
	Han Fang\footnote{Corresponding authors.}\textsuperscript{\rm 3},
	Zehua Ma\textsuperscript{\rm 1}, \\
	Weiming Zhang\footnotemark[1]\textsuperscript{\rm 1},
	Nenghai Yu\textsuperscript{\rm 1}\\
}
\affiliations{
	\textsuperscript{\rm 1}University of Science and Technology of China,\\
	\textsuperscript{\rm 2}University of Waterloo,\\
	\textsuperscript{\rm 3}National University of Singapore\\
	hichangliu@mail.ustc.edu.cn, jiezhangsp@gmail.com, fanghan@nus.edu.sg,\\ mzh045@mail.ustc.edu.cn, zhangwm@ustc.edu.cn, ynh@ustc.edu.cn\\
	
}

\begin{document}
	\maketitle
	
	\begin{abstract}
		Audio watermarking is widely used for leaking source tracing. The robustness of the watermark determines the traceability of the algorithm. With the development of digital technology, audio re-recording (AR) has become an efficient and covert means to steal secrets. AR process could drastically destroy the watermark signal while preserving the original information. This puts forward a new requirement for audio watermarking at this stage, that is, to be robust to AR distortions. Unfortunately, none of the existing algorithms can effectively resist AR attacks due to the complexity of the AR process. To address this limitation, this paper proposes DeAR, a \underline{de}ep-learning-based \underline{a}udio \underline{r}e-recording resistant watermarking. Inspired by DNN-based image watermarking, we pioneer a deep learning framework for audio carriers, based on which the watermark signal can be effectively embedded and extracted.
		Meanwhile, in order to resist the AR attack, we delicately analyze the distortions that occurred in the AR process and design the corresponding distortion layer to cooperate with the proposed watermarking framework. 
		Extensive experiments show that the proposed algorithm can resist not only common electronic channel distortions but also AR distortions. Under the premise of high-quality embedding (SNR=$25.86\,\mathrm{dB}$), in the case of a common re-recording distance ($20\,\mathrm{cm}$), the algorithm can effectively achieve an average bit recovery accuracy of $98.55$\%. 
	\end{abstract}
	
	\section{Introduction}
	As an efficient method to trace the source of leakage, digital watermarking technology has been widely studied. There have been many excellent works in image \cite{fang2018screen}, audio \cite{erfani2016audio}, and video \cite{asikuzzaman2017overview} watermarking. 
	The two most important properties that audio watermarking should satisfy are fidelity and robustness. Fidelity ensures the normal use of the watermarked audio. Robustness guarantees that even if the audio is distorted (MPEG encoding, noise addition, audio re-recording, etc.), the embedded watermark can still be extracted losslessly\cite{wu1999robust}.

\begin{figure}[t] 		
		\centering	
		\includegraphics[scale=0.5]{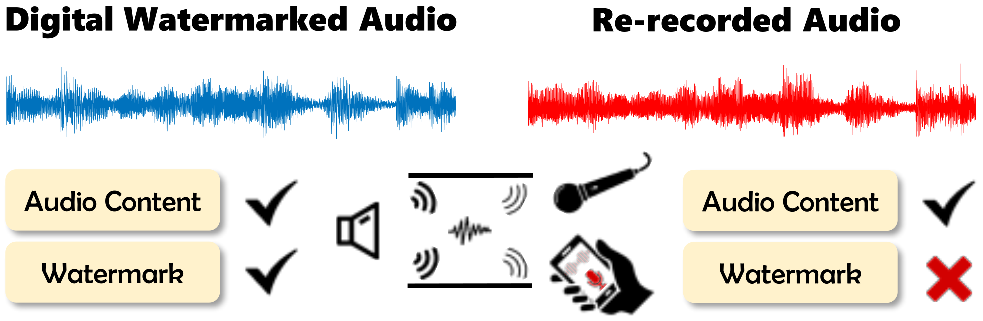}
		\caption{The re-recording operation preserves the content information while destroying the watermark information in the audio to hide the source of the leakage.}
		\label{fig:AR}
	\end{figure}

	Most traditional audio watermarking methods are concerned with the robustness of digital distortions in the electronic channel because most audio copying occurs in the digital channel. However, with the miniaturization of recording devices, audio re-recording (AR) has become a more convenient and converted way to copy audios. Meanwhile, AR could effectively preserve the audio content and significantly damage the embedded watermark signal; attackers can easily and stealthily steal the audio content without any watermark evidence, as shown in \Fref{fig:AR}. Therefore, ensuring robustness to AR becomes the urgent property of audio watermarking at this stage.

	Currently, the research field of audio watermarking is still dominated by traditional mathematical algorithms, which try to search for an invariant feature before and after distortion to conduct watermark embedding.
	Most used features are in the transform domain, such as discrete cosine transform (DCT), discrete wavelet transform (DWT), and fast Fourier transform (FFT) \cite{su2018snr, xiang2017spread, natgunanathan2012robust, liu2018patchwork}. However, due to the complexity of the AR process itself, quantitatively and delicately analyzing the distortions and finding the robust feature in this process is a nontrivial task. Therefore, none of the existing algorithms can resist AR distortion very well. This drives the demand to seek a learnable feature that is adapted to AR distortions.

	In recent years, deep-learning-based watermarking has achieved tremendous success in the field of image and video watermarking \cite{zhu2018hidden,tancik2020stegastamp,zhang2019robust,luo2021dvmark}. The main backbone of such a framework is an end-to-end auto-encoder-like architecture that contains an encoder, a distortion layer, and a decoder. The central part to ensure robustness is the distortion layer which tries to generate the distorted image for training \cite{zhu2018hidden}, and it is essential to make the distortion layer differentiable. However, the AR process is a complex and non-differential process. Although some algorithms currently focus on deep learning-based information hiding studies in the audio field \cite{kreuk2020hide,jiang2020smartsteganogaphy}, they cannot handle the robustness requirement against distortions, including AR distortion.

	To address the robustness of AR distortion, in this paper, we propose DeAR, a \underline{de}ep-learning-based \underline{a}udio \underline{r}e-recording resistant watermarking method, which can effectively resist AR distortion at different distances in the real world.
	Specifically, we adopt the classic deep-learning-based watermarking framework, where an encoder and a decoder are jointly trained for watermark embedding and extraction.
	To achieve robustness against re-recording, we first analyze the re-recording process from the effects of sound propagation in the air 
	and the processing of microphones and speakers.
	According to the analysis, we delicately model the AR distortion with several differential operations (environment reverberation, band-pass filtering, and Gaussian noise) and serve such operations as the distortion layer to cooperate with the proposed framework.
	In addition, motivated by traditional audio watermarking, we change the input of the time domain signal into the low-frequency coefficients of the audio for better robustness. Specifically, 
	we introduce the differential time-frequency transform and its corresponding reverse transform into the end-to-end training process, which can automatically search for the optimal embedding frequency rather than be determined by a pre-defined rule such as Singular Value Decomposition (SVD). 
	Experimental results verify that the proposed method can achieve satisfying robustness against audio re-recording at different distances and is also resilient to other common distortions.

	The primary contributions of our work are concluded as follows:
	\begin{itemize}
		\item
		We are the first to propose a deep-learning-based audio watermarking against audio re-recording (AR), DeAR, which can achieve watermark embedding and extraction in an end-to-end manner rather than based on handcrafted rules. According to the audio cover, we flexibly adapt audio frequency transform and one-dimensional convolution operations for better implementation.
		\item
		To achieve robustness against AR, we first analyze the distortions induced by AR and model it as a distortion pipeline composed of environment reverberation, band-pass filtering, and Gaussian noise. The whole pipeline is further absorbed into the training of DeAR.
		\item
		Extensive experiments demonstrate that the proposed method can achieve robustness against audio re-recording and common electronic channel distortions while guaranteeing the requirement of fidelity. In addition, some ablation studies further verify our design and prove the flexibility of DeAR.
	\end{itemize}
	
	\section{Related Work} \label{sec:rw}
	
	\subsection{Traditional Audio Watermarking}
	
	Traditional audio watermarking (AW) mainly embeds watermark information in the time domain and the transform domain. Time-domain based AW \cite{cvejic2004increasing, natgunanathan2010novel} is simple and efficient but not robust enough. In comparison, transform-domain based AW \cite{bansal2015comparative,su2018snr} can achieve better robustness. 
	For example, Su \etal\cite{su2018snr} conducted DWT transform and DCT transform on the audio and appended the watermark information in the low-frequency band to achieve robustness. However, its watermark extraction is non-blind, requiring the access to the raw audio. Besides, they did not consider robustness against re-recording process. 
	Based on the above method, Liu \el\cite{liu2018patchwork} first considered the re-recording robustness in traditional framework. Specifically, they simplify the practical process as an idealized noise appending process, which only influences the amplitude of the target audio. However, Liu's method does not perform well in a practical scenario.
	We explained that re-recording is composed of not only the additive noise but also other different distortions such as the convolutional noise (namely, environment reverberation) and noise induced by microphones and speakers, which is demonstrated in many existing works \cite{peddinti2015reverberation, yakura2018robust}.

	\subsection{Real-world Air Channel Distortion}
	In this paper, we regard the re-recording process as an air channel distortion.
	Eliminating the side effect of the air channel distortion has always been an inspiring topic in automatic speech recognition (ASR) tasks. To address this issue, massive perturbed data are collected for training noise adaptive acoustic models, which is not applicable for the audio watermarking task. Except that, the most similar research field is robust adversarial audio of speech-to-text models, which guarantees robustness against air channel distortion. In detail, the corresponding studies \cite{qin2019imperceptible,yakura2018robust} modeled these distortions by expectation over transformation (EOT) operations, which is incorporated in the optimization procedure of the adversarial audio. Motivated by this, we first analyze the re-recording process from the effects of sound propagation in the air and the processing of microphones and speakers. Then, we model the re-recording as a distortion pipeline, which is composed of environment reverberation, band-pass filtering, and Gaussian noise. This distortion pipeline can easily cooperate with the deep-learning-based audio watermarking framework. 
	
	\subsection{Deep-learning-based Image Watermarking}
	
\begin{figure}[t]
		\centering
		\includegraphics[scale=0.48]{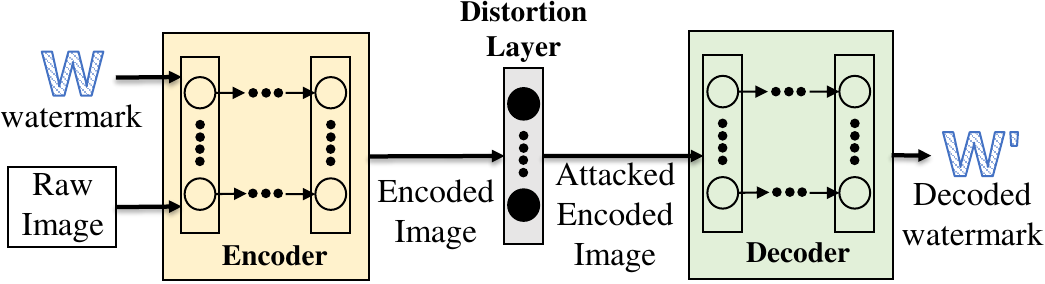}
		\caption{Illustration of the deep-learning-based image watermarking framework.}
		\label{basic_framework}
	\end{figure}

	In recent years, some deep-learning (DL)-based image watermarking schemes have been proposed. For example, Zhu \el \cite{zhu2018hidden} proposed the first end-to-end image watermarking framework. As shown in \Fref{basic_framework}, the popular framework contains three parts: an encoder responsible for watermark embedding, a differential distortion layer to simulate the transmission process, and a decoder structure responsible for watermark extraction. The main challenge is making the non-differential distortion differential, which is necessary for end-to-end training. 
	
	In \cite{zhu2018hidden}, Zhu \el leveraged a differential approximations
	for non-differential JPEG compression, which was further improved by subsequent works \cite{ahmadi2020redmark,ying2021hiding}. Based on \cite{zhu2018hidden}, Tancik \el \cite{tancik2020stegastamp} further considered the distortion caused by printing and camera-shooting, Jia 
	\cite{jia2020rihoop} realized resilience to 3D rendering, and Zhang \el \cite{zhang2020model} even achieved robustness against model extraction. 
	
	Based on DL-based image watermarking, there have been a few attempts at DL-based audio steganography. For example, Jiang \etal \cite{jiang2020smartsteganogaphy} converted raw audio to 2D mel-spectrum images by short-time Fourier transform (STFT) to satisfy the above image watermarking framework. However, STFT and its inverse transform can intrinsically induce the loss of embedded information. In addition, this method lacks the consideration of any robustness during transmission, including audio re-recording (AR).
	
	\begin{figure*}[t]
		\centering
		\includegraphics[scale=0.565]{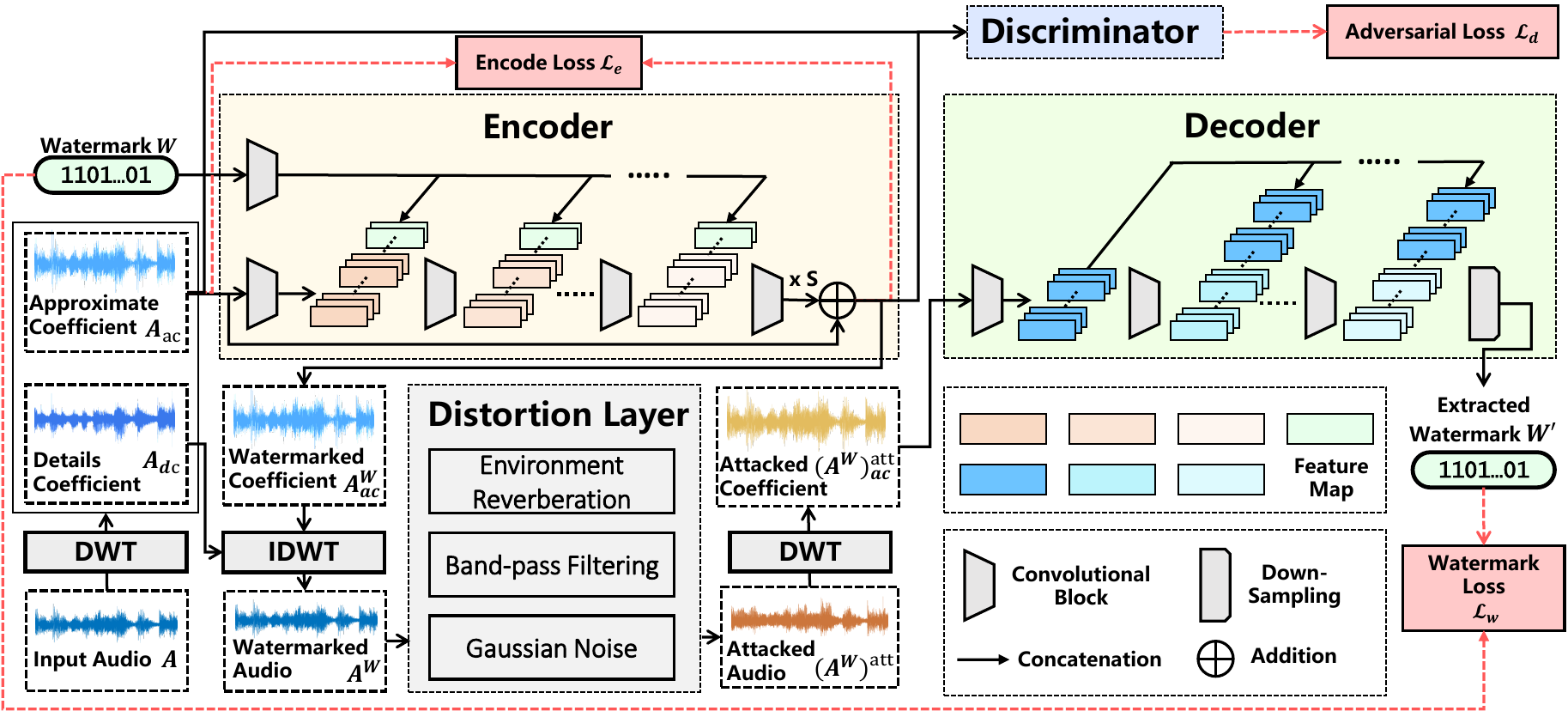}
		\caption{The overall framework of the proposed DeAR.}
	\label{framework_1}
\end{figure*}

\section{Method} \label{sec:pm}

As shown in \Fref{framework_1}, DeAR is mainly composed of three parts, namely, an encoder for watermark embedding, a decoder for watermark extraction, and a distortion layer to enhance robustness against audio re-recording. All of the above parts are jointly trained, and we introduce each part in detail.

\subsection{Watermark Embedding}
We use $A$ to represent a single-channel raw audio with length $N$. Rather than directly feeding the raw audio $A$ into the encoder, we first transfer it into the frequency domain by a differential DWT and obtain the corresponding approximate coefficients $A_{ac}$ and detail coefficients $A_{dc}$, \ie,
\begin{linenomath*}
	\begin{equation}
		A_{ac}, A_{dc} = \operatorname{DWT}(A),
	\end{equation}
\end{linenomath*}
where the length of $A_{ac}$ and $A_{dc}$ is half the original audio signal, namely, $N/2$.
Motivated by traditional audio watermarking, we propose embedding the watermark into the low frequency of the raw audio, namely, leveraging $A_{ac}$ as the audio cover. Meanwhile, $A_{dc}$ is deposited for subsequent audio reconstruction. The goal of the encoder is to embed the watermark information $W$ into $A_{ac}$. As shown in \Fref{framework_1}, we want the encoder $\mathbf{En}$ to generate a stealthy residual $\mathcal{R}$ and further stamp it onto the original $A_{ac}$ to generate the watermarked approximate coefficients $A_{ac}^W$, \ie,
\begin{linenomath*}
	\begin{equation}
		A_{ac}^W = \mathbf{En}(A_{ac},W)*S + A_{ac},
	\end{equation}
\end{linenomath*}
where $S$ is the strength factor and set as 1 by default. To satisfy the fidelity requirement, we constrain the watermarked approximate coefficients $A_{ac}^W$ acoustically consistent with the original one $A_{ac}$. To achieve this, we introduce a basic loss $\mathcal{L}_{e}$ for the encoder's training, that is, we adopt the widely-used mean square error MSE as $\mathcal{L}_{e}$, \ie,
\begin{linenomath*}
	\begin{equation}
		\begin{aligned}
			\mathcal{L}_{e} &= \operatorname{MSE}(A_{ac},A_{ac}^W)
			&= \frac{2}{N}\sum_{i=1}^{N/2}(A_{ac}(i)-A_{ac}^W(i))^2.
		\end{aligned}
	\end{equation}
\end{linenomath*}
To further improve the fidelity and minimize the domain gap between $A_{ac}^W$ and $A_{ac}$, we introduce an extra discriminator $\mathbf{D}$ for adversarial training with the encoder $\mathbf{En}$, and adversarial loss $\mathcal{L}_{d}$ will let $\mathbf{En}$ embed watermarks better so that $\mathbf{D}$ cannot distinguish $A_{ac}^W$ from the watermark-free $A_{ac}$, \ie,
\begin{linenomath*}
	\begin{equation}
		\mathcal{L}_{d} =  log(1-\mathbf{D}(A_{ac}^W)).
	\end{equation}
\end{linenomath*}
Meanwhile, for $\mathbf{D}$, $\mathcal{L}_{D} = log(1-\mathbf{D}(A_{ac}))+log(\mathbf{D}(A_{ac}^W))$.

\subsection{Watermark Extraction}
Given the watermarked approximate coefficients $A_{ac}^W$, the decoder $\mathbf{De}$ needs to recover watermark $W^{\prime}$ as consistent as the original watermark $W$.
To achieve this, we introduce the watermark loss $\mathcal{L}_w$, namely, the MSE between the original watermark $W$ and the extracted watermark $W^{\prime}$, \ie,
\begin{linenomath*}
	\begin{equation}
		\mathcal{L}_w = \operatorname{MSE}(W,W^{\prime}).
	\end{equation}
\end{linenomath*}
It should be emphasized that we adopt the binary watermark $W\in \{-1,1\}^{L}$ instead of $\{0,1\}^{L}$, which is better for forensics. 
Meanwhile, this helps the MSE-based constraint work better.

\subsection{Audio Re-recording Modeling}
To enhance the robustness against the audio re-recording (AR) process, we further insert a distortion layer between the encoder and the decoder. As mentioned above, it is essential to make the distortion layer differentiable, which can prevent gradient interruption during end-to-end learning. However, the AR process is a complex and non-differential process. To overcome this challenge, we learn from the studies on real-world air channel distortion \cite{qin2019imperceptible,yakura2018robust}, and introduce a differential audio re-recording operation $\mathbf{DAR}$, which consists of three components: environment reverberation, band-pass filtering, and Gaussian noise. Because $\mathbf{DAR}$ is a processing pipeline operated on the time domain, it cannot be directly applied to $A_{ac}^W$. Therefore, after generating $A_{ac}^W$, we adopt inverse DWT (IDWT) to transform the watermarked approximate coefficients $A_{ac}^W$ back to the watermarked audio $A^W$, with the corresponding deposited $A_{dc}$, \ie, 
\begin{linenomath*}
	\begin{equation}
		A^W = \operatorname{IDWT}(A_{ac}^W, A_{dc}).
	\end{equation}
\end{linenomath*}

\subsubsection{Environment Reverberation.}
The impulse Response (IR) is the environment's reaction when presented with a brief input signal. It describes the acoustic characteristics of an environment, particularly of interest being the behavior of reverberation in the space. IR can reproduce the reverberation in the captured environment by convolution. From different microphones, room environments, and speakers, we can collect diverse base IR $p$ to form a set $P$. Given a target audio $A$, we randomly select a base IR $p$ from the dataset $P$, and operate convolution $Conv(\cdot)$ on $A$ by $p$ to simulate the Environment Reverberation (ER). \ie,

\begin{linenomath*}
	\begin{equation}
		\operatorname{ER(A)} =  Conv(A, p), \quad\text{where $p \in P$}.
	\end{equation}
\end{linenomath*}
Here, we follow the previous work \cite{palomaki2004techniques} and leverage the acoustic impulse responses dataset\footnote{https://www1.icsi.berkeley.edu/Speech/papers/gelbart-ms/pointers/} that was collected using four mics and various degrees of reverberation in the varechoic chamber at Bell Labs. 

\subsubsection{Band-pass Filtering.}
Since the frequency band of human hearing is limited, the widely-used normal range is from $500$ Hz to $2000$ Hz. Based on this, commonly-used speakers will not play audio with a too high or too low-frequency band. Meanwhile, the microphone will also process the playing audio, usually cutting off the frequency band outside the normal range to reduce noise, that is, a basic denoising process. Therefore, to simulate the distortions caused by the inherent characteristics of devices such as the speaker and the microphone, we apply a frequency band-pass filtering BF($\cdot$) operation to the watermarked audio. Given a target audio $A$, we conduct BF($\cdot$) as follows:
\begin{linenomath*}
	\begin{equation} \label{equ:bp}
		\operatorname{BF}(A) = {LF}[{HF}[A, \operatorname{\alpha}], \operatorname{\beta}],
	\end{equation}
\end{linenomath*}
where $LH[\cdot]$ and $HF[\cdot]$ represent the low-pass filtering and the high-pass filtering, respectively. And $\alpha$ and $\beta$ denote the corresponding thresholds of $HF[\cdot]$ and $LF[\cdot]$.

\subsubsection{Gaussian Noise.}
In addition to the above two components, we introduce the popular Gaussian noise to simulate the random noise induced by indeterminate factors during AR process. Gaussian noise is a kind of additive noise, that is widely used in current automatic speech recognition (ASR) systems \cite{yin2015noisy} to enhance the robustness against random environmental noise. Specifically, we operate GN($\cdot$) on audio $A$ by directly superimposing noise $\omega\sim\mathcal{N}(0, \sigma^2)$, \ie,
\begin{linenomath*}\begin{equation}
		\begin{split}
			\operatorname{GN}(A) = A + \omega,
			\text{where $\omega\sim\mathcal{N}(0, \sigma^2)$}.
		\end{split}
	\end{equation}
\end{linenomath*}
It shall be mentioned that the $\sigma$ is audio-aware and determined by the pre-defined signal-noise-ratio, which is randomly sampled from $20\,\mathrm{dB}$ to $25\,\mathrm{dB}$.

\subsubsection{The pipeline of the distortion layer.}
$\mathbf{DAR}$ is set as the distortion layer to enhance robustness against AR process. The pipeline of $\mathbf{DAR}$ is as follows:
\begin{linenomath*}
	\begin{equation}
		\mathbf{DAR}(\cdot) = \operatorname{GN}(\operatorname{BF}(\operatorname{ER}(\cdot))).
	\end{equation}
\end{linenomath*}
Given a watermarked audio $A^W$, we can finally obtain the attacked audio $(A^W)^{att}$, \ie,
\begin{linenomath*}
	\begin{equation}
		(A^W)^{att} = \mathbf{DAR}(A^W).
	\end{equation}
\end{linenomath*}
Afterward, we leverage DWT to acquire its corresponding approximate coefficients $(A^W)^{att}_{ac}$ and feed it into the decoder $\mathbf{De}$ for watermark extraction, \ie,
\begin{linenomath*}
	\begin{equation}
		\begin{aligned}
			(A^W)^{att}_{ac},(A^W)^{att}_{dc} &= \operatorname{DWT}((A^W)^{att}),\\
			W^{\prime} &= \mathbf{De}((A^W)^{att}_{ac}).
		\end{aligned}
	\end{equation}
\end{linenomath*}

\subsection{More Details of DeAR}

\subsubsection{Network Structures.}
For the encoder $\mathbf{En}$, we adopt a fully convolutional network, which keeps the size of the feature maps in each layer unchanged. Similar to $\mathbf{En}$,  the decoder $\mathbf{De}$ and the discriminator $\mathbf{D}$ also utilize a fully convolutional network but further append convolutional downsampling module and mean operation for final extraction and binary classification, respectively. To remedy the information loss during forward propagation, we leverage the skip connection to stack the 
initial information with the feature map in each layer for both $\mathbf{En}$ and $\mathbf{De}$. More importantly, we leverage 1D convolution rather than 2D convolution in basic networks, which is effective for the 1D audio waveform.

\subsubsection{Loss functions.}
During the training stage, we jointly train the encoder and the decoder, and the whole loss function $\mathcal{L}$ can be formulated as follows: 
\begin{linenomath*}
	\begin{equation}
		\mathcal{L} = \lambda_e\mathcal{L}_{e} + \lambda_d\mathcal{L}_{d} + \lambda_w\mathcal{L}_{w},
	\end{equation}
\end{linenomath*}
where $\lambda_e$, $\lambda_d$, and $\lambda_w$ are used to balance the three terms.

\section{Experiments}
\subsection{Experiment Settings}

\subsubsection{Dataset.} We conduct our experiments on FMA \cite{fma_dataset}, a famous music analysis dataset in which $12000$ audios are utilized for the training of the proposed DeAR, and $200$ randomly selected audios are adopted as testing audios. The sampling frequencies are all 44.1kHz and are processed as wav audio with 500k sample points in length.

\subsubsection{Metrics.}
To measure the fidelity of the watermarked audio, the signal-to-noise ratio (\textbf{SNR}) is adopted with the following definition:
\begin{equation*}
	\rm{SNR} = 10\cdot \log\left(\frac{\sum_{i=1}^NA(i)^2}{\sum_{i=1}^N\left[A^W(i)-A(i)\right]^2}\right).
\end{equation*}
Besides, we take the average bit recovery accuracy $\overline{\textbf{ACC}}$, calculated from $200$ testing audios, to evaluate the robustness of watermarking schemes.

\subsubsection{Implementation Details.}
In the training process of DeAR, we set $\lambda_e=150$, $\lambda_w=1$ and $\lambda_d=0.01$, and utilize Adam\cite{kingma2014adam} with a learning rate of $10^{-4}$ for optimization by default. And we empirically set the threshold of the high-pass filtering ($HF[\cdot]$) and that of the low-pass filtering ($LF[\cdot]$), namely, $\alpha$ and $\beta$ as $1\,\mathrm{kHz}$ and $4\,\mathrm{kHz}$ for training the model robust to re-recording.

In the testing process, the same watermark bit sequence of $100$ bits is embedded for all testing audios. We adopt methods the most relevant to the proposed DeAR as the baseline methods, \ie, Liu's method \cite{liu2018patchwork} and Su's method \cite{su2018snr}. For a fair comparison, we slightly modify the parameter of Su's method for better robustness\footnote{The upper bound of the interval of the search space of intensity factor is set to $2$.}. 

\begin{figure}[t]
	\centering
	\includegraphics[scale=0.25]{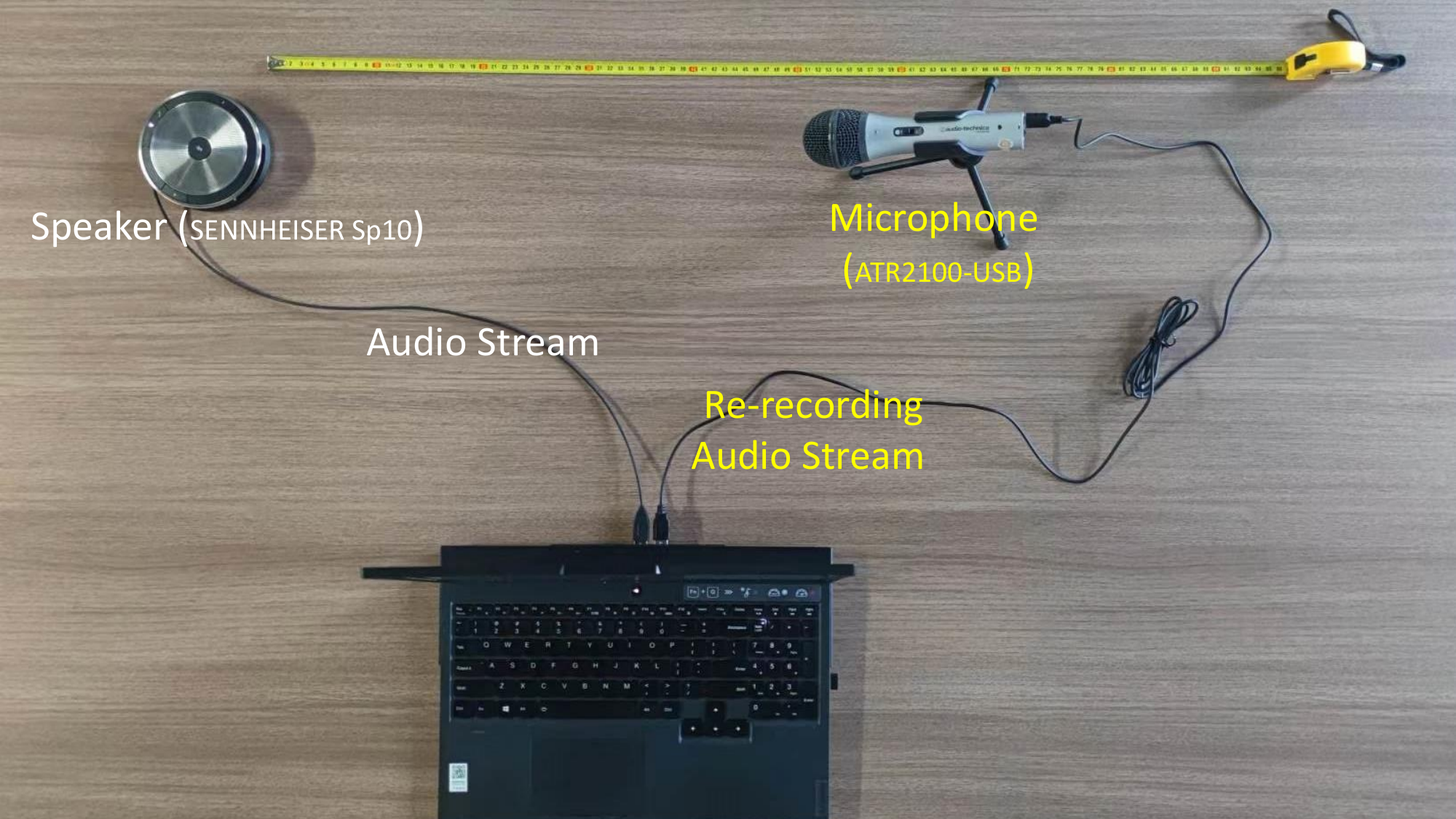}
	\caption{A practical experimental environment to evaluate robustness against audio re-recording.}
	\label{series}
\end{figure}

For re-recording experiments, we use one consumer-grade speaker, SENNHEISER Sp10, as the playback device and a consumer-grade microphone, ATR2100-USB, to re-record the played audio. The corresponding experimental scenario is shown in \Fref{series}, where $5\,\mathrm{cm}$ is set as the default distance between the speaker and the microphone. Considering the de-synchronization introduced by digital-to-analog conversion, we design a synchronization strategy and apply it to DeAR and all baseline methods for a fair comparison. Specifically, we shift the target re-recorded audio within a pre-defined range $(3/44.1\,\mathrm{s}, 8/44.1\,\mathrm{s})$ and calculate the corresponding bit recovery accuracy (\textbf{ACC}), in which the highest one is regarded as the final \textbf{ACC}.

\subsection{Comparison Results}   

\begin{figure*}[t]
	\begin{center}
		\includegraphics[scale=0.103] {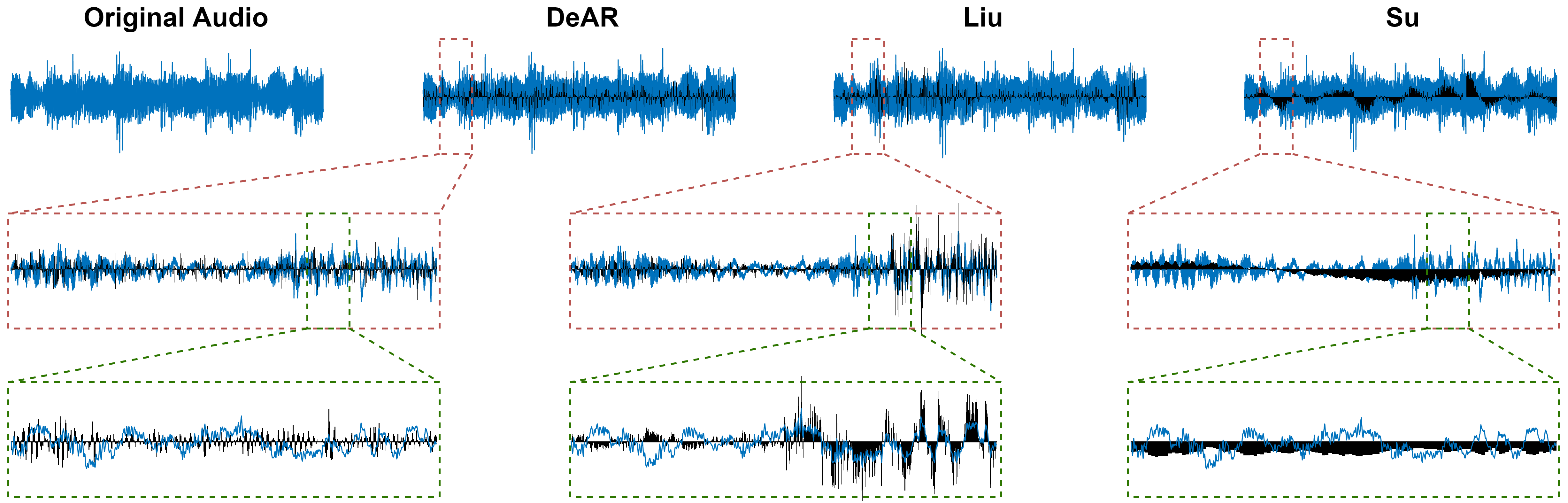}
		\caption{Qualitative comparison of fidelity. The top row shows the original audio and watermarked audios of DeAR and the baseline methods, wherein the black component represents the 10$\times$ residuals. To better illustrate their difference, the enlarged views of the same regions are shown in the middle and bottom rows.
		}
		\label{fig:waveform}
	\end{center}
\end{figure*}

\begin{table}[t]
	\begin{center} 
		\setlength\tabcolsep{16pt}
		\begin{tabular}{c|c c c}
			\toprule[1.5pt]
			Metrics   &  DeAR      & Liu     		& Su 	\\ \hline
			SNR ($\mathrm{dB}$)      & \bv{25.86}         & 25.81    & 24.94   \\
			$\overline{\textbf{ACC}}$ ($\%$)		  & \bv{99.18}			 & 77.09	  & 56.00 \\
			\bottomrule[1.5pt]
		\end{tabular}
		\caption{Quantitative comparison with the baseline methods.}
		\label{tab:cmp}
	\end{center}
\end{table}

\begin{table}[t]
	\setlength\tabcolsep{9pt}
	\begin{tabular}{ccccc}
		\toprule[1.5pt]
		Distance ($\mathrm{cm}$) & 5              & 20             & 50            & 100            \\ \hline
		DeAR         & \textbf{99.18} & \textbf{98.55} & \textbf{93.40} & \textbf{92.68} \\
		Liu          & 77.09          & 82.64          & 74.76         & 66.02         \\
		\bottomrule[1.5pt]
	\end{tabular}
	\caption{Comparison of robustness against re-recording at different distances.}
	\label{tab:dis}
\end{table}

\begin{figure}[t]
	\begin{center}
		\includegraphics[scale=0.1045] {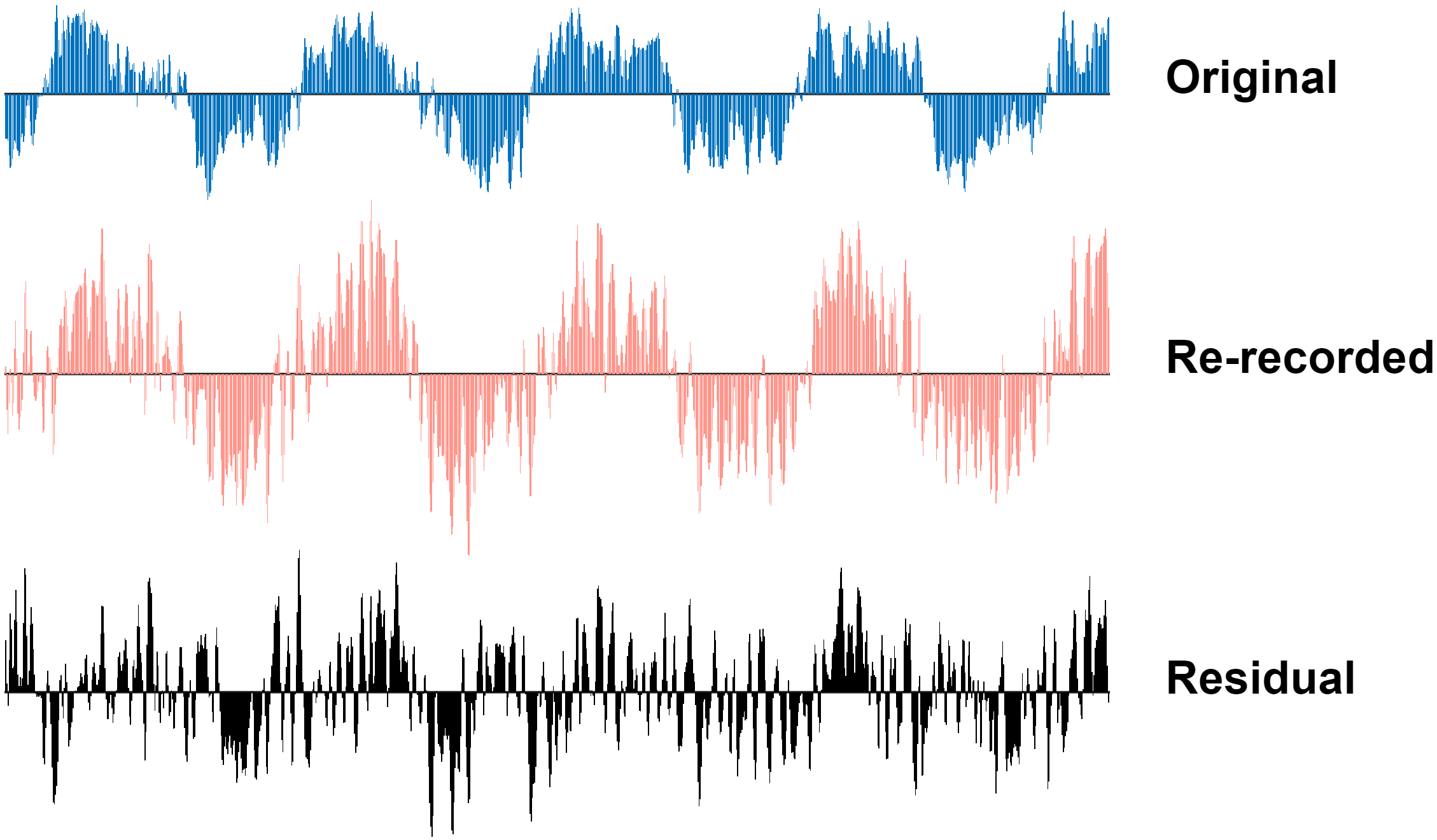}
		\caption{Visual example of noise induced by audio re-recording.}
		\label{fig:res}
	\end{center}
\end{figure}

\subsubsection{Fidelity.}   

We first compare the fidelity of the proposed DeAR with the baseline methods. As shown in \Tref{tab:cmp}, DeAR achieves a 25.86 SNR, which outperforms the baseline methods (SNR is 25.81 and 24.94 for Liu's method \cite{liu2018patchwork} and Su's methods \cite{su2018snr}, respectively). Furthermore, we provide one visual example of the watermarked audio for qualitative comparison in \Fref{fig:waveform}. 
We can observe that the proposed DeAR and Liu's method tend to modify the original audio adaptively, while Su's method prefers a relatively large and uniform noise pattern. Compared with Liu's method, DeAR induces more slight modification to guarantee fidelity.

\subsubsection{Robustness Against Audio Re-recording.}    
We compare robustness against audio re-recording in this experiment and provide quantitative results in \Tref{tab:cmp}. With comparable fidelity, DeAR outperforms the baseline methods by a large margin (above $20$\% and $40$\%, respectively). Su's method is fragile to the audio re-recording (AR) process because it only considers the common distortions during the digital transmission but lacks the consideration of AR. For Liu's method, we believe that its limited performance ($77.09$\%) is due to its AR model being an idealized additive noise, which is inconsistent with the practical scenario. In \Fref{fig:res}, we provide a visual example of the noise induced by the practical re-recording process, which is a more complex noise pattern with an even larger amplitude than the original audio.
In addition to the default distance ($5\,\mathrm{cm}$), we further conduct a controlled comparison with Liu's method at different distances. As shown in \Tref{tab:dis}, our method performs better over a wide range of distances. As the distance increases, the robustness against AR suffers a corresponding degradation but is still acceptable (all above $90$\%).

\begin{table}[t]
	\begin{center}
		\centering
		\setlength\tabcolsep{5.2pt}
		\begin{tabular}{ c | c | c  c  c }
			\toprule[1.5pt]
			\multicolumn{2}{c|}{Distortions}     & DeAR                & Su             & Liu                     \\
			\hline
			\multirow{ 4}{*}{\tabincell{c}{Gaussian\\Noise}}
			& $10\,\mathrm{dB}$                         & 87.13 / 98.67                        & \bv{100.0} & 60.09 \\
			
			& $15\,\mathrm{dB}$                         & 93.77 / 99.89                        & \bv{100.0} & 62.41 \\
			
			& $20\,\mathrm{dB}$                         & 96.10 / 99.99                        & \bv{100.0} & 65.76 \\
			
			& $25\,\mathrm{dB}$                         & 96.96 / 99.99                        & \bv{100.0} & 71.61 \\
			\hline
			\multirow{2}{*}{MP3}
			& $64\,\mathrm{kbps}$                       & 95.22 / \bv{99.94} 				 & 97.98                        & 88.54 \\
			& $128\,\mathrm{kbps}$                      & 97.00 / \bv{99.97} 				 & 98.00                        & 98.53 \\

			\hline
			\multirow{ 2}{*}{Band-pass}
			& $1\,\mathrm{kHz}$                   & 98.17 / \bv{100.0} 				 & \underline{57.02}                        & 99.14 \\
			& $4\,\mathrm{kHz}$                     & 92.12 / 99.06                       & \bv{99.99} & \underline{50.57}\\

			\hline
			\multicolumn{2}{c|}{Re-sampling}   & 97.04 / \bv{100.0} 				 & \bv{100.0} & 99.37                                \\
			\hline
			\multicolumn{2}{c|}{Dropout}       & 97.14 / \bv{99.99} 				 & 99.08                        & 69.63               \\
			\hline
			\multicolumn{2}{c|}{Amplitude Modification} & 97.14 / \bv{99.99} 				 & 98.00                        & 99.84               \\
			\hline
			\multicolumn{2}{c|}{Re-quantization}    & 97.11 / \bv{100.0} 				 & \bv{100.0} & 94.22                                \\
			\hline
			\multicolumn{2}{c|}{Median Filtering }   & 96.16 / \bv{100.0} 				 & \bv{100.0} & 90.80                                \\
			\bottomrule[1.5pt]
		\end{tabular}
		\caption{Robustness against other common distortions. We provide the $\overline{\textbf{ACC}}$ of the default / the enhanced DeAR. }
		\label{tab:common}
	\end{center}
\end{table}

\begin{figure*}[ht]
	\begin{center}
		\includegraphics[scale=0.340] {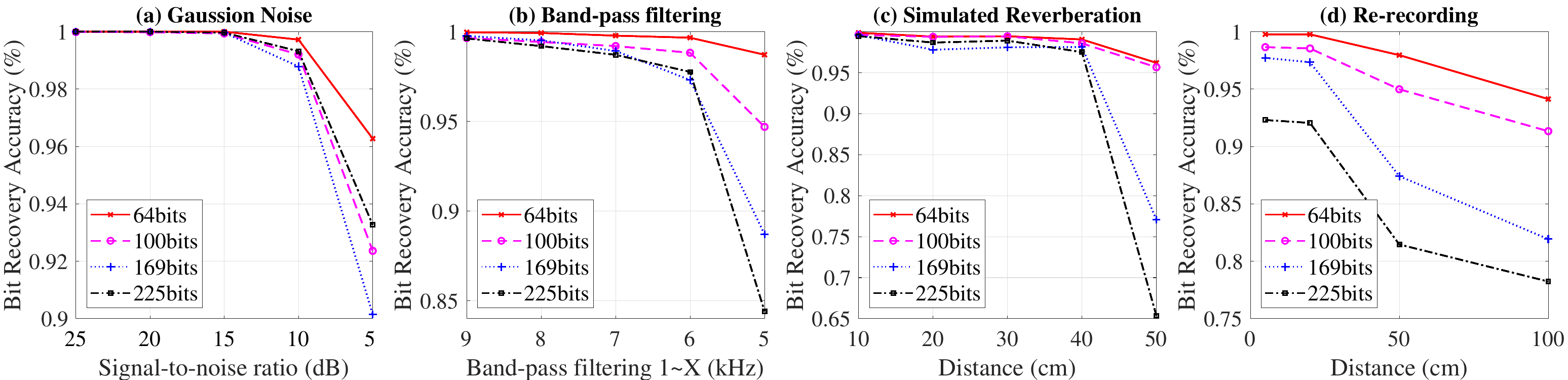}
		\caption{The influence of different embedding bits on robustness. }
		\label{line-chart}
	\end{center}
\end{figure*}

\begin{table}[t]
	\setlength\tabcolsep{9pt}
	\begin{tabular}{ccccc}
		\toprule[1.5pt]
		Distance ($\mathrm{cm}$) & 5              & 20             & 50            & 100            \\ \hline
		Default       & \textbf{99.18} & \textbf{98.55} & 93.40 & \textbf{92.68} \\
		Enhanced        & 98.65         & 98.54          & \textbf{94.98}      & 91.33         \\
		\bottomrule[1.5pt]
	\end{tabular}
	\caption{Comparison of robustness against AR between the default and the enhanced DeAR. }
	\label{tab:enhance}
\end{table}

\subsubsection{Robustness Against Other Common Distortions.}
To compare robustness more comprehensively, we further evaluate it under other common distortions during digital transmission, namely,  Gaussian noise under different signal-to-noise ratios ($10\,\mathrm{dB}$, $15\,\mathrm{dB}$, $20\,\mathrm{dB}$, $25\,\mathrm{dB}$), MP3 compression ($64\,\mathrm{kbps}$, $128\,\mathrm{kbps}$),  Band-pass ($1\,\mathrm{kHz}$ high-pass, $4\,\mathrm{kHz}$ low-pass), Re-sampling (the watermarked audios are resampled to 90\% of the original sampling frequency and then resampled back), Dropout (drop (zero) a sample every 100 sample points), Amplitude Modification (90\% of the original audios), Re-quantization (requantize down to 8 bits/sample), and Median Filtering (window size of 3).
As shown in \Tref{tab:common}, Su's method achieves excellent robustness in most cases except under the $1\,\mathrm{kHz}$ high-pass (underlined data), which may be the reason for its fragility to audio re-recording. Liu's method cannot extract the watermark well, facing $4\,\mathrm{kHz}$ low-pass. In contrast, DeAR is robust under all types of distortions.

Furthermore, we propose the enhanced DeAR by adding additional distortions to the distortion layer in addition to the AR distortion. And one distortion is randomly selected for each mini-batch during training. The additional distortions contains the above common distortions except Gaussian noise and band-pass filtering. For example, it only involves Gaussian noise under $10\,\mathrm{dB}$ for the enhanced training. As shown in \Tref{tab:common} and \Tref{tab:enhance}, the enhanced DeAR can achieve better robustness against other common distortions while still preserving robustness against audio re-recording. 

\subsection{Ablation Study}

\subsubsection{The Influence of Different Embedding Bits.}
We further explore the influence of different embedding bits on robustness. We take $64\,$bits, $100\,$bits, $169\,$bits, and $225\,$bits as examples to evaluate the robustness against Gaussian noise, band-pass, simulated reverberation, and audio re-recording. We evaluate the robustness of the model with different embedding bits under the exact SNR requirement ($SNR=26\pm0.2$). For each distortion, we also test on different strengths. We shall note that the simulated reverberation is implemented by convolving watermarked audios and impulse responses. The impulse response here is selected in a conference room at different distances between the speaker and the microphone.
As shown in \Fref{line-chart}, in most cases, we shall sacrifice the robustness ability to some extent if we want to embed more information into the target audio.

\subsubsection{The Importance of Each Component of the Re-recording Modeling.}
\mzh{To fully verify the importance of the proposed re-recording modeling, we re-train the model with the modified distortion layer three times. In each training, one component of three distortions in the distortion layer is removed.} Besides the default consumer-grade microphone, we also utilize the widely-used smartphone (\eg Apple iPhone12 pro) to re-record the watermarked audio. As shown in \Tref{component}, \mzh{all designed distortion} components, \mzh{\ie}, environment reverberation, band-pass filtering, and Gaussian noise could improve the robustness against audio re-recording process. Among them, \mzh{the} reverberation is much necessary \mzh{because it induces an accuracy improvement of more than 25\%.}

\begin{table}[t]
	\setlength\tabcolsep{6.5pt}
	\begin{center}
		\begin{tabular}{c|cccc}
			\toprule[1.5pt]
			Device     	& Default           	& w/o ER 	& w/o BF         	& w/o GN \\ \hline
			Microphone 	& \textbf{99.18}    	& 73.63       	& 98.52 		 	& 75.80            \\
			Smartphone	& \textbf{98.57} 		& 66.17			& 92.64          	& 75.25           \\
			\bottomrule[1.5pt]
		\end{tabular}
	\end{center}
	
	\caption{The robustness performance ($\overline{\textbf{ACC}}$) of different configurations under different devices.}
	\label{component}
\end{table}

\subsubsection{Flexibility with Strength Factor. }
Given an audio coefficient and watermark as input, the well-trained encoder outputs a watermark residual, which is further superimposed on the input. In a practical scenario, we are able to flexibly adjust the strength of the residual to balance the trade-off between fidelity and robustness. In \Tref{tab-snr}, we provide the quantitative results of fidelity and robustness against audio re-recording under different strength factors, which demonstrate the flexibility of the proposed DeAR.

\begin{table}[t]
	\begin{center}
		\setlength\tabcolsep{5.5pt}
		\begin{tabular}{ c | c  c  c  c  c}
			\toprule[1.5pt]
			Strength Factor   		& 0.2  		& 0.5		& 0.8		& 1			& 1.2 		\\
			\hline
			SNR ($\mathrm{dB}$) 					& 39.84     & 31.88 	& 27.79		& 25.86		& 24.27     \\	
			$\overline{\textbf{ACC}}$ ($\%$)   	 				& 79.58 	& 94.81     & 98.48     & 99.18		& 99.49		\\
			\bottomrule[1.5pt]
		\end{tabular}
		\caption{Performance of DeAR with different strength factors.}
		\label{tab-snr}
	\end{center}
\end{table}

\section{Conclusion}
To the best of our knowledge, we are the first to propose using deep neural networks for audio watermarking and achieve robustness against the audio re-recording (AR) process.
To achieve this, we jointly train an encoder and decoder for watermark embedding and extraction, between which a distortion layer that simulates the re-recording process is further inserted to enhance robustness. Extensive experiments demonstrate that our method outperforms the baseline methods in terms of both fidelity and resilience to the AR process. Furthermore, some ablation studies are also conducted to verify the importance of our design and the flexibility in practical scenarios.

\section{Acknowledgments}
This work was supported in part by the Natural Science Foundation of China under Grant 62072421, 62002334, 62102386, 62121002 and U20B2047 and Students' Innovation and Entrepreneurship Foundation of USTC under Grant XY2022X01CY.

\bibliography{ref}
\end{document}

%% file: command.tex
\usepackage{color}
\definecolor{green}{rgb}{0, 0.5, 0}
\definecolor{orange}{rgb}{0.8, 0.6, 0.2}
\definecolor{orange2}{rgb}{1.0, 0.6, 0.2}
\definecolor{red}{rgb}{1.0, 0.0, 0.0}
\definecolor{teal}{rgb}{0.0, 0.4, 0.4}
\definecolor{purple}{rgb}{0.65,0,0.65}
\definecolor{saffron}{rgb}{0.95,0.75,0.2}
\definecolor{turquoise}{rgb}{0.0,0.5,0.5}
\definecolor{black}{rgb}{0.0, 0.0, 0.0}
\definecolor{gray}{rgb}{0.5, 0.5, 0.5}

\newcommand{\mzh}[1]{{\color{black}#1}}

\newcommand{\etal}{\textit{et al}. }
\newcommand{\el}{\textit{et al}. }
\newcommand{\ie}{\textit{i}.\textit{e}.}
\newcommand{\eg}{\textit{e}.\textit{g}. }

\newcommand{\Tref}[1]{Table~\ref{#1}}

\newcommand{\Fref}[1]{Fig.~\ref{#1}}

\newcommand{\bv}[1]{\textbf{#1}}

\newcommand{\tabincell}[2]{\begin{tabular}{@{}#1@{}}#2\end{tabular}}